\documentclass[aps,prc,preprint,amsmath,showpacs,superscriptaddress,nofootinbib]{revtex4-1}
\usepackage{graphicx,epstopdf,epsfig,psfrag}
\newcommand{\beqy}{\begin{eqnarray}}
\newcommand{\eeqy}{\end{eqnarray}}
\newcommand{\bmlet}{\begin{subequations}}
\newcommand{\emlet}{\end{subequations}}

\begin{document}

\title{Analytical determination of the structure and nuclear abundances of the outer crust of a cold nonaccreted neutron star}

\author{N. Chamel}
\affiliation{Institut d'Astronomie et d'Astrophysique, CP-226, Boulevard du Triomphe, Universit\'e Libre de Bruxelles, 1050 Brussels, Belgium}

\begin{abstract}
A very fast iterative method is presented to calculate the internal constitution of the outer crust of a cold nonaccreted neutron star, making use of very accurate analytical formulas for the transition pressures between adjacent crustal layers and their density. In addition to the composition of the different crustal layers, their depth and their baryonic mass content can be simultaneously estimated using an approximate solution of Einstein's general relativistic equations. The overall computing time is drastically reduced compared to the traditional approach, thus opening the door to large-scale statistical studies and sensitivity analyses. 
\end{abstract}

\keywords{dense matter, neutron star crust, abundance}

\maketitle

\section{Introduction}

Formed in the aftermath of gravitational core-collapse supernova explosions, neutron stars are among the most compact stars in the Universe. A few meters below their solid surface, atoms are fully ionized by the tremendous gravitational pressure: matter thus consists of bare atomic nuclei arranged on a crystal lattice in a charge compensating background of highly degenerate relativistic electrons. The deeper regions are expected to be stratified into different layers (see, e.g. Ref.~\cite{blaschke2018} for a recent review). At some pressure $P=P_{\rm drip}$, neutrons drip out of nuclei thus delimiting the boundary between the outer and inner crusts (see, e.g. Ref.~\cite{chamel2015} for a recent discussion). 

Although the outer crust of a neutron star represents a small fraction of the stellar 
mass, it may be dynamically ripped off by tidal and pressure forces during the collision of two neutron stars, or a neutron star and a black hole. The subsequent decompression  of this neutron-rich material provides suitable conditions for the rapid neutron capture process so called r-process at the origin of stable and some long-lived radioactive neutron-rich nuclides heavier than iron~\cite{arnould2007}. The final nuclear abundances depend on the initial composition of the neutron-star crust~\cite{goriely2011,goriely2011b}. This scenario has been recently confirmed by the monitoring of the kilonova following the detection of gravitational waves from the binary neutron-star merger GW170817~\cite{abbott2017}. The analysis of the electromagnetic emission indicates that the entire outer crust was ejected and disseminated in the interstellar medium. 

Since the pioneer studies of Refs.~\cite{tondeur71,bps71}, the composition of the outer crust of a cold nonaccreted neutron star has been numerically determined under the cold-catalyzed matter hypothesis~\cite{hw58,htww65} by minimizing the Gibbs free energy per nucleon $g$ at zero temperature and for a finite set of pressure values (see, e.g. Refs.~\cite{ruester2006,guo2007,roca2008,pearson2011,kreim2013,wolf2013,bcpm,utama2016,pearson2018}). The only input are the masses of all possible nuclei, most of which have not been experimentally measured but can be calculated using various nuclear models~\cite{lunney2003}. As shown in Ref.~\cite{hemp2013}, some crustal layers can be easily missed if the pressure step is not small enough. However, such layers may still represent a sizable fraction of the crustal mass, especially if they lie in the densest regions. A fine enough pressure grid is therefore required to properly calculate nuclear abundances. The computational cost of such calculations can thus become prohibitive, especially for large-scale statistical studies, as recently undertaken in Ref.~\cite{pastore2019}. For the same reason, early studies made use of semi-empirical mass formula and were restricted to a very small subset of nuclei. For instance, only 130 nuclei were considered in the minimization performed in the seminal work of Ref.~\cite{bps71} while about $10^4$ nuclei are expected to exist~\cite{nazarewicz2018}. Moreover, those 130 selected nuclei were all made of even numbers of neutrons and protons. Although even-even nuclei are generally more stable than their neighbors in the nuclear chart due to pairing, the presence of odd nuclei in neutron-star crusts cannot be ruled out a priori since the equilibrium state is also determined by the electron gas and its interactions with ions. As a matter of fact, odd nuclei, such as $^{79}$Cu and $^{121}$Y, have been predicted by some models~\cite{pearson2011,kreim2013,wolf2013}. Early results, which are still popular today (especially those of Ref.~\cite{bps71}), should thus be employed with some care.

In this paper, a very fast and accurate iterative method is presented to calculate analytically the stratification of the outer crust of a cold nonaccreted neutron star.

\section{Transition between adjacent crustal layers}
\label{sec:transition}

In the following, the crustal region at densities $\rho$ above the ionization threshold and below the neutron-drip point will be considered. Although various multinary ionic compounds might be present in the crust of accreted neutron stars (see, e.g. Ref.\cite{chamel2017}), their existence in nonaccreted neutron star is expected to be marginal~\cite{chamel2016}. It is thus assumed that each crustal layer is made of a single nuclear species ($A$, $Z$) with mass number $A$ and atomic number $Z$ in thermodynamic equilibrium at temperatures $T$ below the crystallization temperature $T_m$. Because $T_m$ is typically much lower than the electron Fermi temperature~\cite{fantina2020}, the electron gas is highly degenerate (for all practical purposes, one can thus set $T=0$~K).

The pressure $P_{1\rightarrow2}$ associated with the transition from a crustal layer made of nuclei ($A_1$, $Z_1$) to 
a denser layer made of nuclei ($A_2$, $Z_2$) is determined by the equilibrium condition
\begin{equation}
\label{eq:equilibrium}
g(A_1,Z_1,P_{1\rightarrow2})=g(A_2,Z_2,P_{1\rightarrow2})\, .
\end{equation}
As shown in Ref.~\cite{chamel2016}, this condition can be solved analytically by expanding the Gibbs free energy per nucleon to first order in the fine structure constant $\alpha=e^2/(\hbar c)$ ($e$ being the elementary electric charge, $\hbar$ the Planck-Dirac constant and $c$ the speed of light). Following the same approach but now taking into account electron exchange and charge polarization corrections given in Ref.~\cite{chamel2016b}, $P_{1\rightarrow2}$ 
can be accurately calculated from the solution of the following equation~:
\begin{equation}\label{eq:threshold-condition}
\mu_e\left(1+\frac{\alpha}{2\pi}\right) + C\, \alpha \hbar c n_e^{1/3}F(Z_1,A_1 ; Z_2, A_2) = \mu_e^{1\rightarrow 2}\, ,
\end{equation}
where $\mu_e$ is the electron Fermi energy, $n_e$ is the electron number density, $C$ is the crystal lattice structure constant, 
\begin{equation}\label{eq:def-F}
F(Z_1, A_1 ; Z_2, A_2)\equiv \left(\frac{4}{3}\frac{Z_{1,\textrm{eff}}^{2/3}Z_1}{A_1} - \frac{1}{3}\frac{Z_{1,\textrm{eff}}^{2/3}Z_2}{A_2} -\frac{Z_{2,\textrm{eff}}^{2/3}Z_2}{A_2}\right)
\left(\frac{Z_1}{A_1}-\frac{Z_2}{A_2}\right)^{-1} \, ,
\end{equation}
\begin{equation}\label{eq:muethres}
\mu_e^{1\rightarrow 2}\equiv \biggl[\frac{M^\prime(A_2,Z_2) c^2}{A_2}-\frac{M^\prime(A_1,Z_1)c^2}{A_1}\biggr]\left(\frac{Z_1}{A_1}-\frac{Z_2}{A_2}\right)^{-1} +  m_e c^2\, ,
\end{equation}
$M^\prime(A,Z)$ denoting the mass of the nucleus ($A$,$Z$) and $m_e$ is the electron mass, and 
\begin{equation}
Z_{\textrm{eff}}=Z\left(1+\alpha\frac{12^{4/3}}{35 \pi^{1/3}}b_1(Z) Z^{2/3}\right)^{3/2}\, ,
\end{equation}
\begin{equation}
b_1(Z)= 1 - 1.1866 \,Z^{-0.267} + 0.27\,Z^{-1}\, .
\end{equation}
The singular case $Z_1/A_1=Z_2/A_2$ needs not be considered as it leads to much higher pressures than any other transition (see, e.g., the discussion in Appendix A of Ref.~\cite{chamel2016}). 

Unlike the density $\rho$, the pressure $P$ varies continuously throughout the star. At the interface between the two layers, the pressure is given by 
\begin{equation}
P_{1\rightarrow2}=P_e(n_e)\left(1+\frac{\alpha}{2\pi}\right) + \frac{C}{3}\, \alpha \hbar c Z_{1,\textrm{eff}}^{2/3} n_e^{4/3}\, , 
\end{equation}
where $P_e$ denotes the pressure of an ideal electron Fermi gas (see, e.g. Ref.~\cite{haensel2007} for general expressions). The associated baryon chemical potential $\mu_{1\rightarrow2}$, which coincides with the Gibbs free energy per nucleon, reads 
\begin{equation}
\mu_{1\rightarrow2} = \frac{M^\prime(A_1,Z_1)c^2}{A_1} + \frac{Z_1}{A_1}\bigg[\mu_e\left(1+\frac{\alpha}{2\pi}\right)-m_e c^2+\frac{4}{3}C \alpha \hbar c  n_e^{1/3} Z_{1,\textrm{eff}}^{2/3} \biggr]\, .
\end{equation}
The transition is
generally accompanied by a discontinuous change of the mean nucleon number density: 
\begin{equation}
\bar n_1^{\rm max} = \frac{A_1}{Z_1} n_e\, , 
\end{equation}
\begin{equation}
\bar n_2^{\rm min} = \frac{A_2}{Z_2} n_e \Biggl\{ 1+\frac{1}{3}C \alpha \hbar c n_e^{1/3} (Z_{1,\textrm{eff}}^{2/3}-Z_{2,\textrm{eff}}^{2/3})\biggl[\frac{dP_e}{dn_e}\left(1+\frac{\alpha}{2\pi}\right)\biggr]^{-1} \Biggr\}\, .
\end{equation}

The bottom of the outer crust is marked by the onset of neutron emission by nuclei. Ignoring neutron-band structure effects~\cite{chamel2007}, this transition is determined by the condition $g=m_n c^2$, where 
 $m_n$ is the neutron mass~\cite{chamel2015}. This condition translates into the following equations
\begin{equation}\label{eq:n-drip-mue}
\mu_e\left(1+\frac{\alpha}{2\pi}\right) + \frac{4}{3}C \alpha \hbar c n_e^{1/3} Z_{\textrm{eff}}^{2/3} =  \mu_e^{\rm drip}\, , 
\end{equation}
\begin{equation}\label{eq:muedrip}
\mu_e^{\rm drip}\equiv \frac{-M^\prime(A,Z)c^2+A m_n c^2}{Z} +m_e c^2 \, .
\end{equation}

Equation~(\ref{eq:threshold-condition}) reduces to a quadratic polynomial equation, which can thus be solved analytically for any degree of relativity of the electron gas~\cite{chamel2016}.  Introducing the dimensionless relativity parameter $x_r=\lambda_e(3\pi^2 n_e)^{1/3}=\sqrt{\gamma_e^2-1}$ with the electron Compton wave length $\lambda_e=\hbar/(m_e c)$ and $\gamma_e=\mu_e/(m_e c^2)$, and considering\footnote{The transition from the outermost layer made of $^{56}$Fe to the layer beneath made of $^{62}$Ni, which is completely determined by experimental measurements, corresponds to $\gamma_e^{1\rightarrow2}\approx 1.9$ MeV, see Table~\ref{tab1}.} $\gamma_e^{1\rightarrow2}>1$ the solution reads
\begin{eqnarray}\label{eq:xr-threshold}
 x_r&=&\gamma_e^{1\rightarrow 2} \Biggl\{\left(1+\frac{\alpha}{2\pi}\right)\sqrt{1-\Biggl[\left(1+\frac{\alpha}{2\pi}\right)^2-\tilde F(Z_1, A_1;Z_2,A_2)^2\Biggr]/(\gamma_e^{1\rightarrow2})^{2}}-\tilde F(Z_1, A_1; Z_2,A_2)\Biggr\}\nonumber \\
&& \times \Biggl[\left(1+\frac{\alpha}{2\pi}\right)^2-\tilde F (Z_1, A_1; Z_2,A_2)^2\Biggr]^{-1}\, .
\end{eqnarray}
with $\gamma_e^{1\rightarrow2}\equiv \mu_e^{1\rightarrow2}/(m_e c^2)$, and 
\begin{equation}
\tilde F(Z_1,A_1;Z_2,A_2) \equiv \frac{C \alpha}{(3\pi^2)^{1/3}} F(Z_1,A_1;Z_2,A_2)\, .
\end{equation}
This solution exists only if $\tilde F(Z_1,A_1;Z_2,A_2)\geq -1$. 
In principle, Eq.~(\ref{eq:threshold-condition}) has two positive distinct roots if $\tilde F(Z_1,A_1;Z_2,A_2) > 1$. However, Eq.~(\ref{eq:xr-threshold}) yields the lowest transition pressure, which is given by%~\cite{chamel2016}  
\begin{eqnarray}
\label{eq:exact-threshold-pressure-1>2}
P_{1\rightarrow2}&=&\frac{m_e c^2}{8 \pi^2 \lambda_e^3} \biggl[x_r\left(\frac{2}{3}x_r^2-1\right)\sqrt{1+x_r^2}+\ln(x_r+\sqrt{1+x_r^2})\biggr]\left(1+\frac{\alpha}{2\pi}\right)\nonumber\\
&&+\frac{C \alpha}{3 (3\pi^2)^{4/3}} x_r^4 \frac{m_e c^2}{\lambda_e^3}Z_{1,\textrm{eff}}^{2/3} 
\, ,
\end{eqnarray}
The associated baryon chemical potential is given by 
\begin{equation}\label{eq:exact-threshold-Gibbs-1>2}
\mu_{1\rightarrow2} = \frac{M^\prime(A_1,Z_1)c^2}{A_1} + \frac{Z_1}{A_1} m_e c^2\bigg[ \sqrt{x_r^2+1}\left(1+\frac{\alpha}{2\pi}\right)-1+\frac{4C \alpha}{3 (3\pi^2)^{1/3}}x_r  Z_{1,\textrm{eff}}^{2/3} \biggr]\, .
\end{equation}
The densities of the adjacent crustal layers are
\begin{eqnarray}
\bar n_1^{\rm max} = \frac{A_1}{Z_1} \frac{x_r^3 }{3 \pi^2 \lambda_e^3}\, ,
\end{eqnarray}
\begin{eqnarray}
\bar n_{2}^{\rm min} = \frac{A_2}{Z_2}\frac{Z_1}{A_1}\bar n_1^{\rm max} \Biggl[1+\frac{C \alpha}{(3 \pi^2)^{1/3}} \biggl(Z_{1,\textrm{eff}}^{2/3}-Z_{2,\textrm{eff}}^{2/3}\biggr)
\frac{\sqrt{1+x_r^2}}{x_r}\left(1+\frac{\alpha}{2\pi}\right)^{-1}\Biggr]\, .
\end{eqnarray}

The neutron-drip pressure $P_{\rm drip}$ and density $\bar n_{\rm drip}$ can be readily obtained from the expressions of $P_{1\rightarrow 2}$ and $\bar n_1^{\rm max}$ respectively replacing $\gamma_e^{1\rightarrow2}$ by 
 $\gamma_e^{\rm drip}\equiv \mu_e^{\rm drip}/(m_e c^2)$ and $\tilde F(Z_1,A_1;Z_2,A_2)$ by $(4/3) C \alpha/(3\pi^2)^{1/3} Z_{\textrm{eff}}^{2/3}$.

\section{Global structure and nuclear abundances}
\label{sec:structure}

The determination of the nuclear abundances in the outer crust of a neutron star requires the calculation of the global structure of the star. In hydrostatic equilibrium, Einstein's equations of general relativity reduce to the well-known Tolman-Oppenheimer-Volkoff (TOV) equations~\cite{tolman1939,oppenheimer1939}
\begin{multline}\label{eq:TOV}
	\frac{{\rm d}P(r)}{{\rm d}r} = -\frac{G\, \mathcal{E}(r)\mathcal{M}(r)}{c^2 r^2}
	\biggl[1+\frac{P(r)}{\mathcal{E}(r)}\biggr] \biggl[1+\frac{4\pi P(r)r^3}{c^2\mathcal{M}(r)}\biggr]\biggl[1-\frac{2G\mathcal{M}(r)}{c^2 r}\biggr]^{-1}\, ,
\end{multline}
where $G$ is the gravitational constant, and
\begin{equation}
	\mathcal{M}(r) = \frac{4\pi}{c^2}\int_0^r\mathcal{E}(r')r'^2{\rm d}r'\, .
\end{equation}
Here $\mathcal{E}(r)$ is the mass-energy density of matter at the radial coordinate $r$. The gravitational mass of the star is given by 
$\mathcal{M}(R)$, where $R$ is the radial coordinate at which the pressure vanishes, $P(R)=0$. 

In the outer crust, the mass-energy density is approximately given by the mass density, $\mathcal{E}\approx \rho c^2$, and $P\ll \rho c^2$. Since the mass $\Delta \mathcal{M}$ contained in the outer crust is typically very small, of order $10^{-5} M_\odot$, where $M_\odot$ is the mass of the Sun, the TOV equations can be approximately expressed in a Newtonian form as \cite{pearson2011}
\begin{equation}\label{eq:Newton}
	\frac{{\rm d}P}{{\rm d}z} \approx g_s \rho \,  ,
\end{equation}
where $z$ is the proper depth below the surface defined by 
\begin{equation}
	z(r) = \int_r^R {\rm d}r^\prime \left(1 - \frac{2G\mathcal{M}(r^\prime)}
	{c^2 r^\prime}\right)^{-1/2}\quad ,
\end{equation}
the surface gravity $g_s$ is given by
\begin{equation}
	g_s = \frac{G\mathcal{M}}{R^2}\left(1 - \frac{r_g}{R}\right)^{-1/2}\, ,
\end{equation}
and $r_g=2 G\mathcal{M}/c^2$ is the Schwarzschild radius. 

The baryonic mass of nucleons contained in a crustal layer of inner and outer radii $r_1$ and $r_2$ is given by 
\begin{equation}
	\delta\,M_B = 4\pi \int_{r_1}^{r_2} r^2 {\Phi(r)}^{1/2}\rho(r)dr \, ,
\end{equation}
with the metric function 
\begin{equation}
	\Phi(r) = \left(1 - \frac{2G\mathcal{M}(r)}{c^2r}\right)^{-1} \, .
\end{equation}
Using Eq.~(\ref{eq:Newton}), replacing $\mathcal{M}(r)$ and $r$ by $\mathcal{M}$ and $R$ respectively, the baryonic mass of the layer can be approximately expressed as 
\begin{equation}\label{eq:thin-crust-mass}
	\delta\,M_B \approx \frac{4\pi R^2}{g_s} \left(1-\frac{r_g}{R}\right)\delta P \, ,
\end{equation}
with $\delta P=P(r_1)-P(r_2)$. The nuclear abundance $\xi_i$ of a layer $i$ is defined by the ratio of the baryonic mass $\delta\,M_B$ to that of the outer crust $\Delta M_B$, defined by 
\begin{equation}
	\Delta\,M_B = 4\pi \int_{r_{\rm drip}}^{R} r^2 {\Phi(r)}^{1/2}\rho(r)dr \, ,
\end{equation}
where $r_{\rm drip}$ is the radial coordinate at the neutron-drip transition, defined by $P(r_{\rm drip})=P_{\rm drip}$. Within the approximation~(\ref{eq:thin-crust-mass}), the nuclear abundance of the layer $i$ is independent of the global structure of the star, and is simply given by 
\begin{equation}\label{eq:xi}
	\xi_i=\frac{\delta M_B}{\Delta M_B} = \frac{\delta P}{P_{\rm drip}}\, .
\end{equation}
With this definition, the sum of the abundances of all crustal layers is normalized as
\begin{equation}
	\sum_i \xi_i=1\, .
\end{equation}
Given the relative abundances, the baryonic mass contained in any layer $i$ can be calculated as 
\begin{equation}\label{eq:baryonic-mass}
	\delta\,M_B \approx \xi_i \frac{8\pi R^4 P_{\rm drip}}{r_g c^2} \left(1-\frac{r_g}{R}\right)^{3/2}  \, .
\end{equation}

Using the analytical expression of Ref.~\cite{zdunik2017} for the thickness $\delta r=R-r$,  the depth $z$ at radial coordinate $r$ can be written as 
\begin{equation}\label{eq:z}
	z \approx \frac{\delta r}{\sqrt{1-r_g/R}}  = \frac{\phi R \sqrt{1-r_g/R}}{1-\phi\left(1-r_g/R\right)} \, ,
\end{equation}
where 
\begin{equation}
	\phi=\frac{R}{r_g}\biggl[\left(\frac{\mu(r)}{\mu(R)}\right)^2-1\biggr]\, .
\end{equation}
At the surface of the star, the baryon chemical potential is simply given by the mass $m_0$ per nucleon of $^{56}$Fe: 
\begin{equation}
	\mu(R) = m_0 c^2 \equiv \frac{M^\prime(56,26)c^2}{56} \approx 930.412~{\rm MeV}\, , 
\end{equation}
using the data from the 2016 Atomic Mass Evaluation (AME)~\cite{AME2016I,AME2016II}. 
 The baryon chemical potential at the bottom of the outer crust (neutron-drip transition) is given by 
\begin{equation}
	\mu(r_{\rm drip})=m_n c^2\approx 939.565~{\rm MeV}\, .
\end{equation} 
Therefore, $\phi$ varies from $0$ at $r=R$ to about $0.02 R/r_g \ll 1$ at $r=r_{\rm drip}$. Since $\phi$ is very small, the depth $z$ can be further approximated by 
\begin{equation}\label{eq:z.approx}
	%z(r) \approx z_{\rm drip} \frac{(\mu(r)/\mu(R))^2-1}{(\mu(r_{\rm drip})/\mu(R))^2-1} \, ,
	z(r) \approx z_{\rm drip} \frac{(\mu(r)/(m_0 c^2))^2-1}{(m_n/m_0)^2-1} \, ,
\end{equation}
where $z_{\rm drip}\equiv z(r_{\rm drip})$ is the depth at the bottom of the outer crust, given by 
\begin{equation}\label{eq:zdrip}
	%z_{\rm drip}\approx \frac{R^2}{r_g}\biggl[\left(\frac{\mu(r_{\rm drip})}{\mu(R)}\right)^2-1\biggr]\sqrt{1-\frac{r_g}{R}}\, .
	z_{\rm drip}\approx \frac{R^2}{r_g}\biggl[\left(\frac{m_n}{m_0}\right)^2-1\biggr]\sqrt{1-\frac{r_g}{R}}\, .
\end{equation}

The precision of Eqs.~(\ref{eq:baryonic-mass}) and (\ref{eq:z}) for typical neutron-star masses and radii is a few \% and less than 1\% respectively~\cite{zdunik2017}. These analytical approximations may actually be more accurate than the numerical solution of the full TOV Eqs.~(\ref{eq:TOV}) that is usually obtained using an interpolated equation of state for which density discontinuities between adjacent layers are smoothed out.

\section{Stratification of the outer crust}

The equilibrium composition of an outer crust layer at given pressure $P$ has been traditionally determined by calculating numerically the minimum of the Gibbs free energy per nucleon $g(A,Z,P)$ among all possible nuclides ($A$, $Z$). This procedure is numerically costly because $g$ does not explicitly depend on the pressure $P$, but is given by 
\begin{equation}\label{eq:gibbs}
g(A,Z,P) = \frac{M^\prime(A,Z)c^2}{A} + \frac{Z}{A}\bigg[\mu_e\left(1+\frac{\alpha}{2\pi}\right)-m_e c^2+\frac{4}{3}C \alpha\hbar c  n_e^{1/3} Z_{\textrm{eff}}^{2/3} \biggr]\, .
\end{equation}
For any given pressure $P$, the electron density $n_e$ must first be calculated by solving the following equation 
\begin{equation}\label{eq:pressure}
P=P_e(n_e)\left(1+\frac{\alpha}{2\pi}\right) + C\, \alpha \hbar c Z_{\textrm{eff}}^{2/3} n_e^{4/3}\, .
\end{equation}
Such inversion must be performed for all possible nuclides (of order $10^4$). Moreover, the minimization must be repeated for a sufficiently large number of pressure values until the neutron-drip transition is reached. 

An alternative approach is proposed, based on the following  idea. Given a crustal layer made of nuclide ($A_1$, $Z_1$), the composition of the layer beneath can be found by merely determining the nuclide ($A_2$, $Z_2$) yielding the lowest transition pressure $P_{1\rightarrow 2}$ and such that $\bar n_2^{\rm min} \geq \bar n_1^{\rm max} $, as required by hydrostatic equilibrium~\cite{chamel2016}. Moreover, the transition must be such that 
\begin{equation}
1\leq \gamma_e^{1\rightarrow2}\leq \sqrt{\left(1+\frac{\alpha}{2\pi}\right)^2-\tilde F(Z_1, A_1;Z_2,A_2)^2}\, ,
\end{equation}
so as to ensure that the real solution~(\ref{eq:xr-threshold}) for the relativity parameter $x_r$ exists.  
Starting from $^{56}$Fe at the stellar surface, the sequence of equilibrium nuclides can thus be determined iteratively. Once the composition has been found, the detailed structure of the crust and the nuclear abundances can be readily calculated using the analytical formulas (\ref{eq:exact-threshold-pressure-1>2}) and (\ref{eq:exact-threshold-Gibbs-1>2}) for the pressure and baryon chemical potential at the interface between adjacent layers. As discussed in Refs.~\cite{chamel2016,chamel2016b}, the relative errors in the transition pressures and densities amount to about 0.1\% at most. Higher precision can be easily achieved once the composition is known by solving numerically the equilibrium condition~(\ref{eq:equilibrium}). The whole procedure is computationally extremely fast, since numerical calculations at each pressure are avoided entirely. 

To illustrate the method, the internal constitution of the outer crust of a cold  nonaccreted neutron star has been calculated using experimental data from the 2016 AME~\cite{AME2016I,AME2016II} supplemented with the microscopic nuclear mass table HFB-27 available on the BRUSLIB database~\cite{bruslib}. These masses were obtained from self-consistent deformed Hartree-Fock-Bogoliubov calculations using the Skyrme effective interaction BSk27~\cite{hfb27}. The very recent measurements of copper isotopes~\cite{welker2017} have been also taken into account. Nuclear masses were estimated from tabulated \emph{atomic} masses after subtracting out the electron binding energy using Eq.~(A4) of Ref.~\cite{lunney2003} (in units of MeV): 
\begin{equation}
M^\prime(A,Z)c^2 = M(A,Z)c^2 + 1.44381\times 10^{-5}\,Z^{2.39} +
1.55468\times 10^{-12}\,Z^{5.35}\, .
\end{equation}
The crystal structure constant was taken from Ref.~\cite{baiko2001}, considering that nuclei are arranged in a body-centered cubic lattice~\cite{chamel2016}. Results are summarized in Table~\ref{tab1}. The overall computations took about 0.06 seconds using an Intel Core i7-975 processor. For comparison, the standard approach using about 18000 different pressure values between $P=P_0=9\times 10^{-12}$ MeV~fm$^{-3}$ (ensuring a mass density $\rho$ greater than $10^6$~g~cm$^{-3}$, a sufficient condition for complete ionization and degeneracy of the electron gas~\cite{blaschke2018}) 
and $P=P_{\rm drip}$ with a pressure step $\delta P=10^{-3}P$ (errors thus being of the same order as for the analytical method) took about 37 minutes, i.e. $\approx 4\times 10^4$ times longer (counting only the time spent in the minimization without solving Einstein's equations for determining the abundances and the depths of the different layers). To better assess the precision of the new method,  Eq.~(\ref{eq:equilibrium}) have been solved sirectly using (\ref{eq:gibbs}) and (\ref{eq:pressure}). Because $P$ depends not only on $n_e$ but also on $Z$, $n_e$ varies discontinuously at the interface between two adjacent layers with different proton numbers $Z_1$ and $Z_2$. The electron densities $n_e^1$ and $n_e^2$ of the two layers, as well as the transition pressure $P_{1\rightarrow2}$ can be obtained from the mechanical  equilibrium condition $P(n_e^1,Z_1)=P(n_e^2,Z_2)=P_{1\rightarrow2}$ together with (\ref{eq:equilibrium}). The relative deviations between the essentially exact results and the analytical formulas are indicated in Table~\ref{tab2}. The errors on the pressures and densities can reach 0.25\%, but are in most cases much smaller of order $10^{-3}$~\% or even less. The errors on the baryon chemical potentials do not exceed $6.4\times 10^{-5}$~\%. The depths are determined with an error of $5.8\times 10^{-2}$~\% at most. As expected, the relative abundances being obtained from pressure differences exhibit larger deviations, up to 2\%. However, these deviations remain within the precision of the thin-crust approximation. In view of this detailed analysis, the full minimization has been repeated with a pressure step $\delta P =10^{-5} P$ for a more relevant comparison with the new method. With a number of pressure points $N\approx \log(P_{\rm drip}/P_0)/\log(1+\delta P/P)\approx 10^6$, the  computing time increased to about 59 hours and 28 minutes. To achieve a precision on the transition pressures and densities of order $10^{-3}$~\%, the traditional approach thus requires $\approx 3.6 \times 10^6$ more computing time than the new method. 

As expected, the most abundant elements (hence the most relevant for the r-process nucleosynthesis) are found in the densest and deepest region of the outer crust, where experimental nuclear mass measurements are not available~\cite{wolf2013}. In particular, the most abundant element is $^{120}$Sr representing about 32\% of the crustal mass, even though it is present in a thin layer, whose extent represents only 8.7\% of the depth at the outer crust bottom. Although the shallower layer made of $^{64}$Ni has a similar extent, its contribution to the crustal mass is negligibly small $-$ 0.065\% $-$ because of its much lower density. The baryonic mass of each crustal layer and their absolute depth can be easily calculated for any given neutron star mass $M$ and radius $R$  using Eqs.~(\ref{eq:baryonic-mass}) and (\ref{eq:z.approx}) respectively.

\section{Conclusions}

A computationally very fast method for determining the structure and the composition of the outer crust of a cold nonaccreted neutron star have been presented. Instead of carrying out numerically the full minimization of the Gibbs free energy per nucleon, very accurate analytical formulas for the pressure and baryon chemical potential at the interface between adjacent layers and their density are used to find iteratively the sequence of equilibrium nuclides starting from the stellar surface down to the neutron-drip transition. 
The nuclear abundances and the depth of the different layers can be calculated simultaneously using approximate analytical solutions of Einstein's equations. Results for any neutron star mass and radius can be easily obtained. The new scheme is found to be tremendously faster than the full numerical minimization, and is therefore particularly well-suited for large-scale statistical studies and sensitivity analyses involving computations over a very large set of different nuclear mass tables. 

Such a fast and accurate analytical scheme would also be highly desirable for the inner crust 
of a neutron star, where nuclear clusters coexist with free neutrons in addition to relativistic
electrons. Indeed, full 3D quantum calculations of the inner crust are computationally extremely
 expensive, and for this reason have thus been limited to a few layers in the densest part of the crust considering fixed proton fractions instead of full beta equilibrium (see, e.g. Ref.~\cite{schuetrumpf2019} and references therein). The Wigner-Seitz approximation reduces significantly the computing time but becomes unreliable at densities above about $0.02$~fm$^{-3}$~\cite{pastore2017}.  
 An alternative approach, originally developed for finite nuclei~\cite{dutta1986} and later adapted to neutron-star crusts~\cite{onsi2008,pearson2012,pearson2015}, is to employ the extended Thomas-Fermi method with consistent shell corrections added perturbatively. This semiclassical approach provides a fast and fairly accurate approximation of the Hartree-Fock-Bogoliubov equations~\cite{shelley2020}, thus opening the door to systematic studies of neutron-star crusts, treating consistently both the outer and inner parts.

\begin{table}
	\centering
	\caption{Stratification of the outer crust of a cold nonaccreted neutron star, as obtained using recent experimental data supplemented with the nuclear mass model HFB-27~\cite{hfb27}. In the table are listed: the atomic numbers $Z_1$ and $Z_2$ of adjacent layers, the corresponding mass numbers $A_1$ and $A_2$, the maximum and minimum mean nucleon number densities $\bar{n}_1^{\rm max}$ and $\bar{n}_2^{\rm min}$ at which the nuclides are  present, the transition pressure 
		$P_{1\rightarrow 2}$, the electron Fermi energy $\mu_e^{1\rightarrow 2}$, the baryon chemical potential $\mu_{1\rightarrow 2}$, the relative abundance $\xi_1$ of nuclide ($A_1$, $Z_1$) and its relative depth $z_1/z_{\rm drip}$. Units are MeV for energy and fm for length. See text for details.}
	\label{tab1}
	\vspace{.5cm}
	\begin{tabular}{|cccccccccccc|}
		\hline
		$Z_1$ & $A_1$ & $Z_2$ & $A_2$   &   $x_r$      &      $\bar n_1^{\rm max}$ &    $\bar n_2^{\rm min}$   &    $P_{1\rightarrow 2}$  &   $\mu_e^{1\rightarrow 2}$   &     $\mu_{1\rightarrow 2}$   &      $\xi_1$           &     $z_1/z_{\rm drip}$ \\
		\hline
		26 & 56 & 28 & 62   &   1.57   &      4.92$\times 10^{-9}$ &    5.06$\times 10^{-9}$ &    3.35$\times 10^{-10}$  &   0.966    &     930.6 &      6.93$\times 10^{-7}$ &    0.0207  \\
		28 & 62 & 28 & 64   &   5.01   &      1.63$\times 10^{-7}$ &    1.68$\times 10^{-7}$ &    4.34$\times 10^{-8}$  &   2.50    &     931.3 &      8.92$\times 10^{-5}$ &    0.0985  \\
		28 & 64 & 28 & 66   &   8.42   &      8.01$\times 10^{-7}$ &    8.26$\times 10^{-7}$ &    3.56$\times 10^{-7}$  &   4.16    &     932.0 &      6.47$\times 10^{-4}$ &    0.177  \\
		28 & 66 & 36 & 86   &   8.61   &      8.83$\times 10^{-7}$ &    9.00$\times 10^{-7}$ &    3.89$\times 10^{-7}$  &   6.21    &     932.1 &      6.89$\times 10^{-5}$ &    0.181  \\
		36 & 86 & 34 & 84   &   11.0  &      1.87$\times 10^{-6}$ &    1.93$\times 10^{-6}$ &    1.04$\times 10^{-6}$  &   5.13    &     932.6 &      1.35$\times 10^{-3}$ &    0.234  \\
		34 & 84 & 32 & 82   &   16.8  &      6.83$\times 10^{-6}$ &    7.08$\times 10^{-6}$ &    5.62$\times 10^{-6}$  &   7.84    &     933.7 &      4.83$\times 10^{-4}$ &    0.357  \\
		32 & 82 & 30 & 80   &   22.3  &      1.68$\times 10^{-5}$ &    1.74$\times 10^{-5}$ &    1.78$\times 10^{-5}$  &   10.5   &     934.8 &      2.52$\times 10^{-2}$ &    0.473  \\
		30 & 80 & 28 & 78   &   28.2  &      3.51$\times 10^{-5}$ &    3.66$\times 10^{-5}$ &    4.53$\times 10^{-5}$  &   13.3   &     935.8 &      5.69$\times 10^{-2}$ &    0.591  \\
		28 & 78 & 44 & 126  &   34.7  &      6.85$\times 10^{-5}$ &    7.12$\times 10^{-5}$ &    1.05$\times 10^{-4}$  &   24.4   &     937.0 &      1.23$\times 10^{-1}$ &    0.717  \\
		44 & 126&  42&  124 &   36.8  &      8.37$\times 10^{-5}$ &    8.62$\times 10^{-5}$ &    1.29$\times 10^{-4}$  &   16.9   &     937.3 &      5.15$\times 10^{-2}$ &    0.752  \\
		42 & 124&  40&  122 &   42.1  &      1.29$\times 10^{-4}$ &    1.34$\times 10^{-4}$ &    2.23$\times 10^{-4}$  &   19.4   &     938.2 &      1.92$\times 10^{-1}$ &    0.847 \\
		40 & 122&  38&  120 &   44.7  &      1.60$\times 10^{-4}$ &    1.66$\times 10^{-4}$ &    2.84$\times 10^{-4}$  &   20.7   &     938.6 &      1.27$\times 10^{-1}$ &    0.893 \\
		38 & 120&  38&  122 &   49.8  &      2.29$\times 10^{-4}$ &    2.33$\times 10^{-4}$ &    4.38$\times 10^{-4}$  &   24.2   &     939.4 &      3.19$\times 10^{-1}$ &    0.980 \\
		38 & 122&  38&  124 &   50.9  &      2.49$\times 10^{-4}$ &    2.53$\times 10^{-4}$ &    4.78$\times 10^{-4}$  &   24.7   &     939.5 &      8.22$\times 10^{-2}$ &    0.998 \\
		38 & 124&  $-$ &  $-$   &   51.1  &      2.55$\times 10^{-4}$ &    $-$            &    4.83$\times 10^{-4}$  &   24.8   &  939.6 &      1.12$\times 10^{-2}$ &    1.00  \\
		\hline
	\end{tabular}
\end{table}

\begin{table}
	\centering
	\caption{Precision of the calculated properties of the outer crust of a neutron star, as listed in Table~\ref{tab1}. The relative deviation $\delta q$ (in \%) of a quantity $q$ is calculated 
		as $\delta q=100(q-q_\textrm{exact})/q_\textrm{exact}$, where $q_\textrm{exact}$ is the exact value while $q$ denotes the value calculated using the analytical formulas. Zero means that the deviation lies within the machine precision. See text for details. }
	\label{tab2}
	\vspace{.5cm}
	\begin{tabular}{|ccccccccccc|}
		\hline
		$Z_1$ & $A_1$ & $Z_2$ & $A_2$   &   $x_r$      &      $\bar n_1^{\rm max}$ &    $\bar n_2^{\rm min}$   &    $P_{1\rightarrow 2}$  &    $\mu_{1\rightarrow 2}$   &      $\xi_1$           &     $z_1/z_{\rm drip}$ \\
		\hline
		26 & 56 & 28 & 62   & 2.1$\times 10^{-3}$ & 6.4$\times 10^{-3}$ &  -3.4$\times 10^{-3}$ &  9.5$\times 10^{-3}$  & 6.9$\times 10^{-7}$ &  -1.4$\times 10^{-2}$ &  3.4$\times 10^{-3}$  \\
		28 & 62 & 28 & 64   & 5.1$\times 10^{-12}$& 1.5$\times 10^{-11}$ & 1.5$\times 10^{-11}$ &  2.1$\times 10^{-11}$ & 0 &      -2.3$\times 10^{-2}$ &    0  \\
		28 & 64 & 28 & 66   & 3.6$\times 10^{-8}$ & 1.1$\times 10^{-7}$ & 1.1$\times 10^{-7}$ & 1.5$\times 10^{-7}$  & 7.0$\times 10^{-11}$ & -2.3$\times 10^{-2}$ & 4.0$\times 10^{-8}$ \\
		28 & 66 & 36 & 86   & 3.3$\times 10^{-2}$ & 1.0$\times 10^{-1}$ & 7.0$\times 10^{-2}$ & 1.4$\times 10^{-1}$  & 6.4$\times 10^{-5}$ &  1.6 & 3.6$\times 10^{-2}$  \\
		36 & 86 & 34 & 84   & 1.2$\times 10^{-3}$ & 3.6$\times 10^{-3}$ & 1.1$\times 10^{-2}$ & 4.9$\times 10^{-3}$  & 2.9$\times 10^{-6}$ & -9.7$\times 10^{-2}$ & 1.3$\times 10^{-3}$ \\
		34 & 84 & 32 & 82   & 1.1$\times 10^{-3}$ & 3.4$\times 10^{-3}$ & 1.1$\times 10^{-2}$ & 4.6$\times 10^{-3}$  & 4.0$\times 10^{-6}$ & -1.9$\times 10^{-2}$ & 1.1$\times 10^{-3}$  \\
		32 & 82 & 30 & 80   & 1.1$\times 10^{-3}$ & 3.2$\times 10^{-3}$ & 1.0$\times 10^{-2}$ & 4.3$\times 10^{-3}$  & 4.9$\times 10^{-6}$ & -1.9$\times 10^{-2}$ & 1.1$\times 10^{-3}$   \\
		30 & 80 & 28 & 78   & 1.0$\times 10^{-3}$ & 3.0$\times 10^{-3}$ & 1.0$\times 10^{-2}$ & 4.0$\times 10^{-3}$  & 5.6$\times 10^{-6}$ & -1.9$\times 10^{-2}$ & 9.6$\times 10^{-4}$  \\
		28 & 78 & 44 & 126  & 6.2$\times 10^{-2}$ & 1.9$\times 10^{-1}$ & 1.2$\times 10^{-1}$ & 2.5$\times 10^{-1}$  & 4.0$\times 10^{-4}$ & 4.1$\times 10^{-1}$ &  5.8$\times 10^{-2}$ \\
		44 & 126&  42&  124 & 1.2$\times 10^{-3}$ & 3.6$\times 10^{-3}$ & 1.2$\times 10^{-2}$ & 4.8$\times 10^{-3}$  & 7.9$\times 10^{-6}$ & -1.0 & 1.1$\times 10^{-3}$  \\
		42 & 124&  40&  122 & 1.1$\times 10^{-3}$ & 3.4$\times 10^{-3}$ & 1.2$\times 10^{-2}$ & 4.6$\times 10^{-3}$  & 8.4$\times 10^{-6}$ & -1.9$\times 10^{-2}$ & 1.0$\times 10^{-3}$ \\
		40 & 122&  38&  120 & 1.1$\times 10^{-3}$ & 3.3$\times 10^{-3}$ & 1.1$\times 10^{-2}$ & 4.4$\times 10^{-3}$  & 8.3$\times 10^{-6}$ & -1.0$\times 10^{-1}$ & 9.5$\times 10^{-4}$ \\
		38 & 120&  38&  122 & 2.9$\times 10^{-14}$& 8.3$\times 10^{-14}$& 7.0$\times 10^{-14}$& 7.4$\times 10^{-14}$ & 0                   & -3.1$\times 10^{-2}$ &    0 \\
		38 & 122&  38&  124 & 0                   & 0                   & 2.1$\times 10^{-14}$& 2.0$\times 10^{-13}$ & 0                   & -2.3$\times 10^{-2}$ &    0 \\
		38 & 124&$-$ &  $-$ & 5.8$\times 10^{-3}$ & 1.7$\times 10^{-2}$ &    $-$              & 2.3$\times 10^{-2}$  & $-$                 &  2.0 &  $-$  \\
		\hline
	\end{tabular}
\end{table}

\section*{Acknowledgments}
This work was financially supported by Fonds de la Recherche Scientifique (Belgium) under grant no. IISN 4.4502.19, and by the European Cooperation in Science and Technology Action CA16214.

\end{document}